\documentclass[aps,prc,longbibliography,twocolumn]{revtex4-1}
\usepackage{physics}
\usepackage{amsmath}
\usepackage{amssymb}
\usepackage{graphicx}
\usepackage{float}
\usepackage{array}
\usepackage{verbatim}
\usepackage[margin=1in]{geometry}
\usepackage{color}

\usepackage{subcaption}
\graphicspath{{images/}}

\begin{document}
\title{Time-dependent Hartree-Fock study of quasifission trajectories in reactions forming $^{294}\text{Og}$}
\author{P. McGlynn, C. Simenel}
\affiliation{Department of Fundamental and Theoretical Physics \& Department of Nuclear Physics and Accelerator Applications, Research School of Physics, Australian National University, Canberra ACT 2600}

\begin{abstract}
\textbf{Background:} Fission modes in superheavy nuclei are expected to be impacted by quantum shell effects. Similar shell effects may  be present in  quasifission reactions, acting to hinder the mass equilibration process in heavy-ion collisions.

\textbf{Purpose:} To investigate quasifission mechanisms in five different reactions forming $^{294}$Og as a compound nucleus and compare quasifission trajectories with predicted fission modes.

\textbf{Methods:} The potential energy surface (PES) of $^{294}$Og is calculated using the static Hartree-Fock approach with BCS pairing correlations. 
Quasifission trajectories for central collisions at various energies are studied with the time-dependent Hartree-Fock theory. 

\textbf{Results:} The exit channel strongly depends on initial mass asymmetry and orientation, but it only exhibits small dependences in the reaction energy. 
The $^{48}$Ca$+^{246}$Cf reaction is affected by the PES topology, leading to either fusion or asymmetric fission. 
Spherical shell effects associated with the $Z=50$ magic gap hinder charge and mass equilibrations in $^{126}$Sn$+^{168}$Er, resulting in large total kinetic energies and compact scission configurations. 

\textbf{Conclusions:} Quasifission trajectories can be interpreted in terms of the underlying PES for low excitation energies. 
Future investigations of quasifission with temperature and angular momentum dependent PES could be considered.
\end{abstract}

\maketitle

\section{Introduction}

Fission of atomic nuclei is one of the most complex nuclear reaction processes. 
Numerous experiments have been carried in the past on fission of long-lived actinides induced by neutron and electromagnetic probes. 
The last two decades also saw major experimental efforts to study fission mechanisms away from stability with relativistic beams, multi-nucleon transfer and fusion reactions \cite{andreyev2017,schmidt2018}.
However, some regions of the nuclear charts remain difficult to access for the purpose of fission investigations, such as neutron-rich nuclei relevant to fission recycling in the r-process \cite{goriely2015b}, and super-heavy nuclei (SHN) which could have a super-asymmetric fission mode (or cluster radioactivity) \cite{poenaru2011,zhang2018,santhosh2018,warda2018,matheson2019,ishizuka2020}, although no conclusive experimental evidence of the existence of this mode have been found so far \cite{banerjee2021}.

Despite recent progresses in microscopic approaches \cite{schunck2016}, theoretical description of fission remains also very challenging \cite{bender2020}. 
Many theoretical approaches use potential energy surfaces (PES) as a major ingredient to predict final fragment properties, such as their mass and charge distributions. 
In particular, quantum shell effects in the compound system \cite{gustafsson1971,cwiok1994,brack1972,bernard2023} and in the fragments \cite{wilkins1976,scamps2018,bernard2023} produce fission valleys in the PES that are usually associated with asymmetric fission modes.

Similar shell effects could also affect the formation of fragments in quasifission reactions \cite{itkis2004,zagrebaev2005,nishio2008,kozulin2014,wakhle2014,morjean2017,hinde2018}. 
Quasifission occurs when two heavy-ions collide, fully dissipate their relative kinetic energy, 
and transfer nucleons from the heavy fragment to the lighter one within a few zeptoseconds ($10^{-21}$~s) to a few tens of zeptoseconds \cite{toke1985,durietz2013,simenel2020}  
(see also \cite{hinde2021} for a recent experimental review on quasifission).
This slow mass drift towards symmetry could eventually be stopped by shell effects in the fragments.

Naturally, there is no guarantee that the resulting quasifission modes are the same as in fission. 
Experimentally, properties of quasifission fragments can be obtained by comparing reactions forming similar compound nuclei with different entrance channels.
Indeed, the most asymmetric ones are expected to have less quasifission~\cite{chizhov2003,rafiei2008,williams2013}. 
Asymmetric modes could then be observed in quasifission while being absent (or with negligible yield) in fusion-fission reactions  \cite{chizhov2003}.
Nevertheless, recent time-dependent Hartree-Fock (TDHF) calculations predicted strong similarities between quasifission fragments produced in $^{50}$Ca$+^{176}$Yb and fission fragments of the $^{226}$Th compound nucleus \cite{simenel2021}, indicating that 
both mechanisms are affected by similar shell effects. 

In principle, quasifission reactions could then be used to explore regions of the PES, including those relevant to fission. 
In particular, one can use the fact that different entrance channels (forming the same compound nucleus) lead to different ``entry points'' on the PES (configurations obtained just after full dissipation of the initial kinetic energy) to search, through quasifission, for different  fission valleys. 
The PES used in fission modelling, however, are often constructed using approximations that do not hold, in principle, in the case of quasifission. 
These approximations include, e.g., axial symmetry, zero angular momentum, and zero temperature (although PES at finite temperatures are sometimes considered \cite{mcdonnell2014,zhao2019b}).
It is not clear, then, if the evolution of a system undergoing quasifission can be interpreted in terms of the topology (barriers, valleys...) of PES used in fission studies.

Here, theoretical predictions of fission and quasifission modes in reactions forming $^{294}$Og as a compound nucleus are presented. 
The properties of the PES of $^{294}$Og are investigated in section~\ref{sec_PES}. Static and time-dependent calculations were performed in the framework of the Skyrme-Hartree-Fock theory with BCS pairing interaction, as discussed in section~\ref{sec_TDHF}. In section~\ref{sec_fission} the fission potential energy surface is analyzed. Simulations of  $^{48}$Ca$+^{246}$Cf, $^{86}$Kr$+^{208}$Pb and $^{126}$Sn$+^{168}$Er heavy-ion collisions leading to quasifission are presented in section~\ref{sec_QF}. Entry points and quasifission trajectories on the PES are studied  and compared to the fission modes in section~\ref{sec_trajs}. Conclusions are drawn in section~\ref{sec_conc}.

\section{Methods \& numerical details}

\subsection{Potential energy surface}
\label{sec_PES}

The potential energy surface (PES) was calculated using the static Hartree-Fock+BCS code SkyAx \cite{reinhard2021}. 
The code solves the Hartree-Fock equations with BCS pairing correlations and constraints on multipole moments.
Axial symmetry is imposed to increase computational efficiency.
The spatial grid used in these calculations has a 1~fm spacing in both the radial and the axial coordinates $r$ and $z$, and spans $-64.5\text{ fm}\leq z\leq 64.5\text{ fm},r\leq 40\text{ fm}$. 

All calculations were performed with the SLy4$d$ parametrisation of the Skyrme  energy-density functional \cite{kim1997} with density-dependent BCS pairing.
Each point in the PES were computed by applying constraints on quadrupole moment $Q_{20}$ and octupole moment  $Q_{30}$ (or on $Q_{20}$ only in computing the adiabatic path) and solving to find the minimal energy configuration satisfying those constraints. The constraints themselves were applied via a damped Lagrange multiplier approach which forces the converged state to have a specific expectation value for each constrained multipole moment. 

Such calculations were performed for a grid of $Q_{20}$ and $Q_{30}$ values ranging from 0 to 270~b and from 0 to 110~b$^{3/2}$ with a spacing of 2~b and 2~b$^{3/2}$, respectively. In some regions of the PES the grid is denser as multiple points were calculated to improve convergence, this is done prescriptively in regions where convergence is difficult, such as near the scission line \cite{dubray2008}. 
The scission line itself is defined from the density of the neck between the fragments.
Here, we consider that scission occurs when the neck density is below $0.08$~nucleons/fm$^3$, i.e., approximatively half the nuclear saturation density. 

\begin{figure}
	\includegraphics[width=\columnwidth]{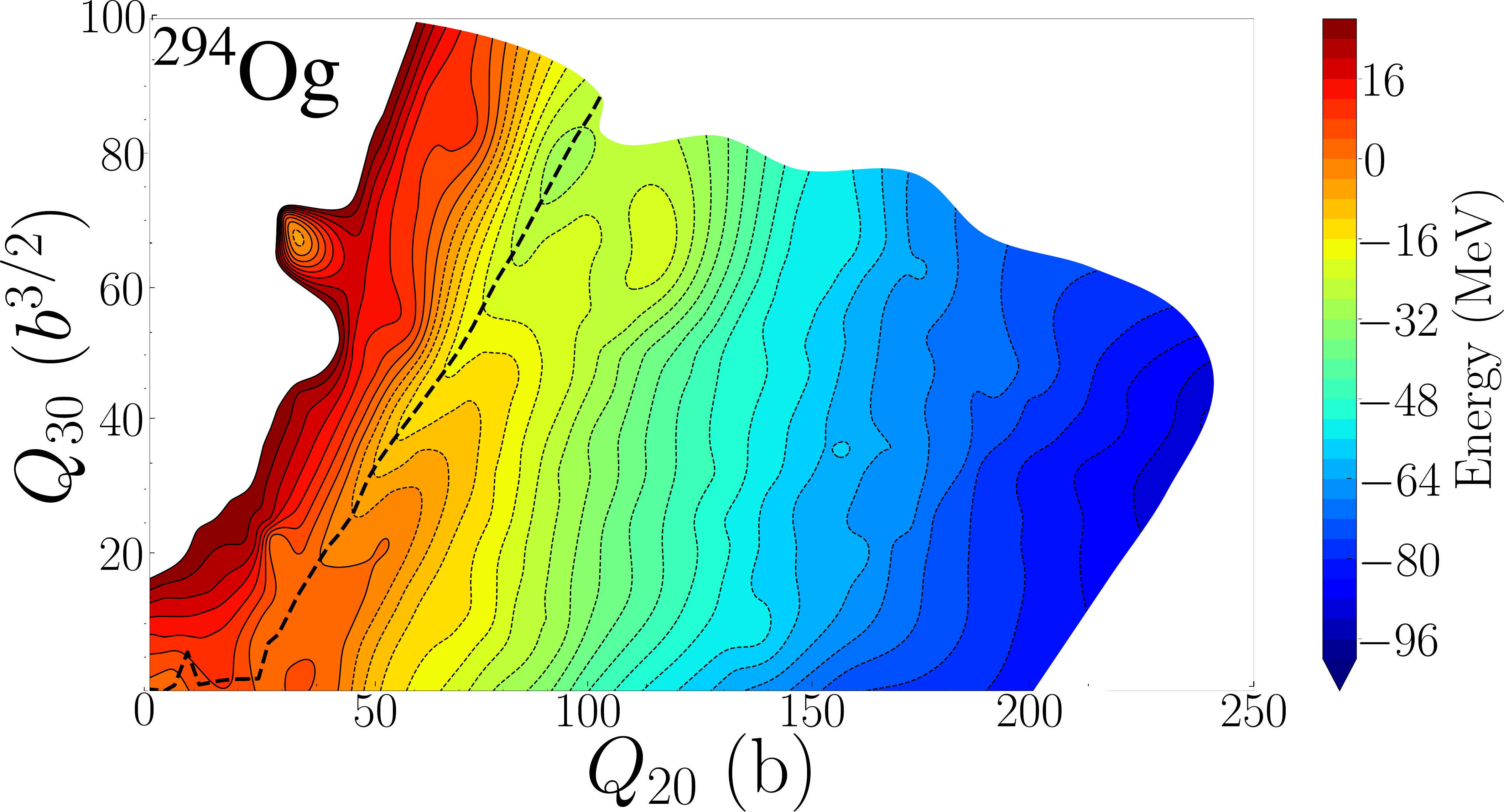}
	\caption{\label{fig_PES}Zero-temperature potential energy surface of oganesson-294. 
	Energy $E$ is relative to the ground-state energy of -2071.4 MeV. The dashed line corresponds to the asymmetric adiabatic path to scission.}

\end{figure}

The resulting PES of $^{294}$Og is shown in Fig.~\ref{fig_PES}. It was generated by interpolation of all the converged points with a radial basis function technique. 
The asymmetric one-dimensional path (dashed line) was determined by performing successive calculations with constraints in $Q_{20}$ (leaving $Q_{30}$ free) and increasing the value of $Q_{20}$ by step of 1~b.

\subsection{Time-dependent Hartree-Fock simulations}
\label{sec_TDHF}

Quasifission mechanisms in reactions forming $^{294}$Og compound nucleus were investigated with the TDHF code  Sky3D \cite{maruhn2014,schuetrumpf2018}. 
A three-dimensional cartesian grid with 1~fm mesh size was used. No-spatial symmetries were assumed. 
The Skyrme energy-density functional and pairing functional are the same as in the PES calculations. 

The collision partners were assumed to be in their ground-state in the initial configuration.
The ground-states were obtained by solving the static Hartree-Fock equations with BCS pairing correlations inside a 28$\times$28$\times$28~fm$^3$ box.
The nuclei were then placed in a larger box of 84$\times$28$\times$28~fm$^3$ with an initial distance of 56~fm between their centres of mass. 
A galilean boost was then applied on each nucleus with opposite linear momenta  and zero angular momentum. 
When one nucleus is initially axially deformed, different orientations of the deformation axis with respect to the collision axis were considered. 

\begin{figure}
	\includegraphics[width=0.6\columnwidth]{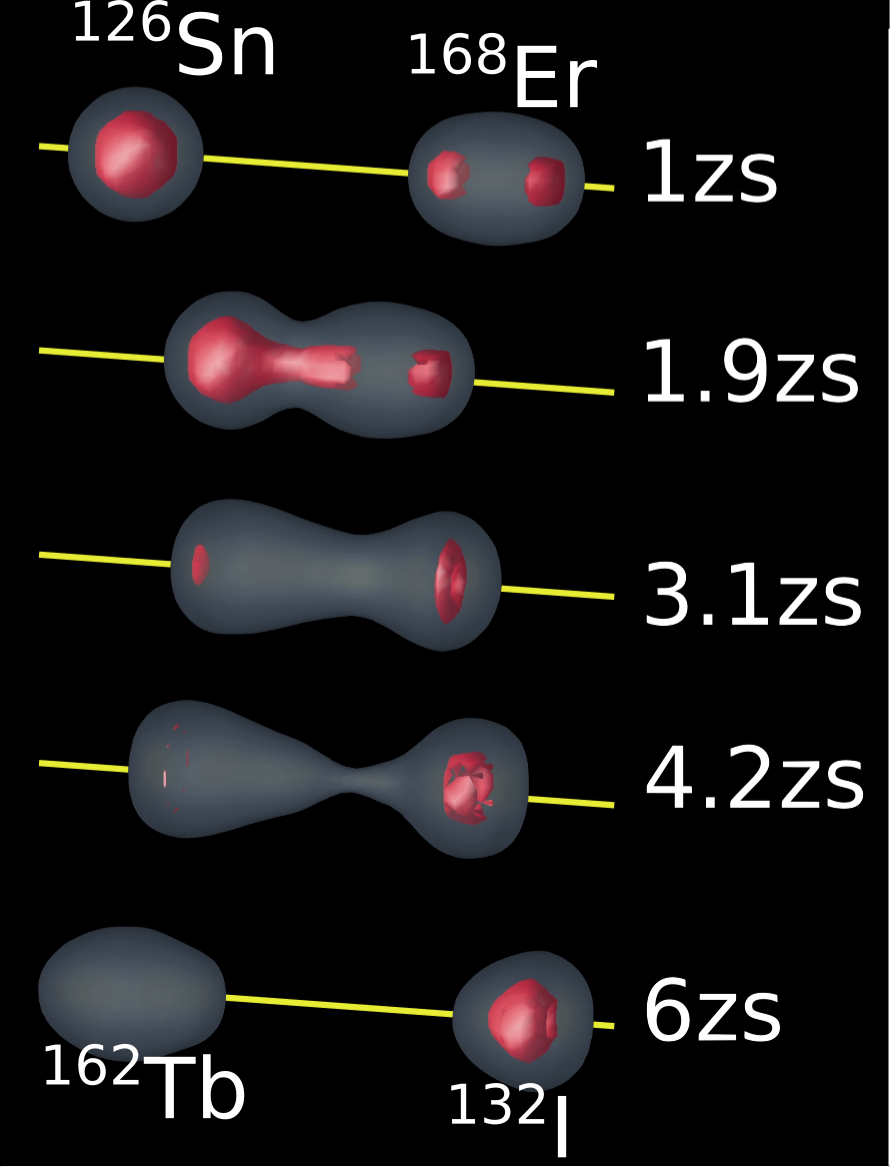}
	\caption{\label{fig_movieframes} Isodensity surfaces at half nuclear saturation $\rho_0/2=0.08$~fm$^{-3}$ (translucent grey) and at saturation density $\rho_0=0.16$~fm$^{-3}$(opaque red), as computed with Sky3D for a quasifission reaction $^{126}$Sn$+^{168}$Er at a centre of mass energy $E_{c.m.}=416$~MeV.  The Collision axis is represented by the yellow solid line. The final fragments are determined based on the average numbers of protons and neutrons.}
\end{figure}

The TDHF equation was solved iteratively with a time step of 0.2~fm/c.
The frozen occupation approximation was used, i.e., the initial single-particle occupation numbers issued by the static BCS calculations were kept constant in time. 
See \cite{negele1982,simenel2012,simenel2018,sekizawa2019,stevenson2019,godbey2020} for reviews of TDHF calculations applied to nuclear dynamics.
An example of evolution of isodensity surfaces in $^{126}$Sn$+^{168}$Er at a centre of mass energy $E_{c.m.}=416$~MeV is given in Fig.~\ref{fig_movieframes}.

The density in the neck between the fragments was used to determine the contact time, defined as the duration of neck density exceeding half the nuclear saturation density ($\rho_0/2=0.08$~fm$^{-3}$). 
The neck location was identified as the minimum density along the collision axis and located between the two largest maxima. 
After  complete reseparation, the sum of the Coulomb interaction energy between the fragments and their kinetic energies becomes constant and was used to define the total kinetic energy (TKE) \cite{simenel2014a} (see also \cite{scamps2015a,tanimura2017,li2023} for similar calculations of TKE in fission).
The number of protons and neutrons in the primary fragments are then determined from integration of neutron and proton densities on each side of the neck.

\section{Fission modes}
\label{sec_fission}

The PES shown in Fig.~\ref{fig_PES} exhibits several structures affecting fission properties of $^{294}$Og.
A clear valley around the adiabatic asymmetric fission path (dashed line) is observed.
It starts just before the second barrier (located at $Q_{20}\simeq35$~b) and goes all the way to scission with the formation of a heavy fragment near the doubly magic $^{208}$Pb nucleus, 
in agreement with earlier studies \cite{poenaru2011,zhang2018,santhosh2018,warda2018,matheson2019,ishizuka2020}.

\begin{figure}
	\includegraphics[width=\columnwidth]{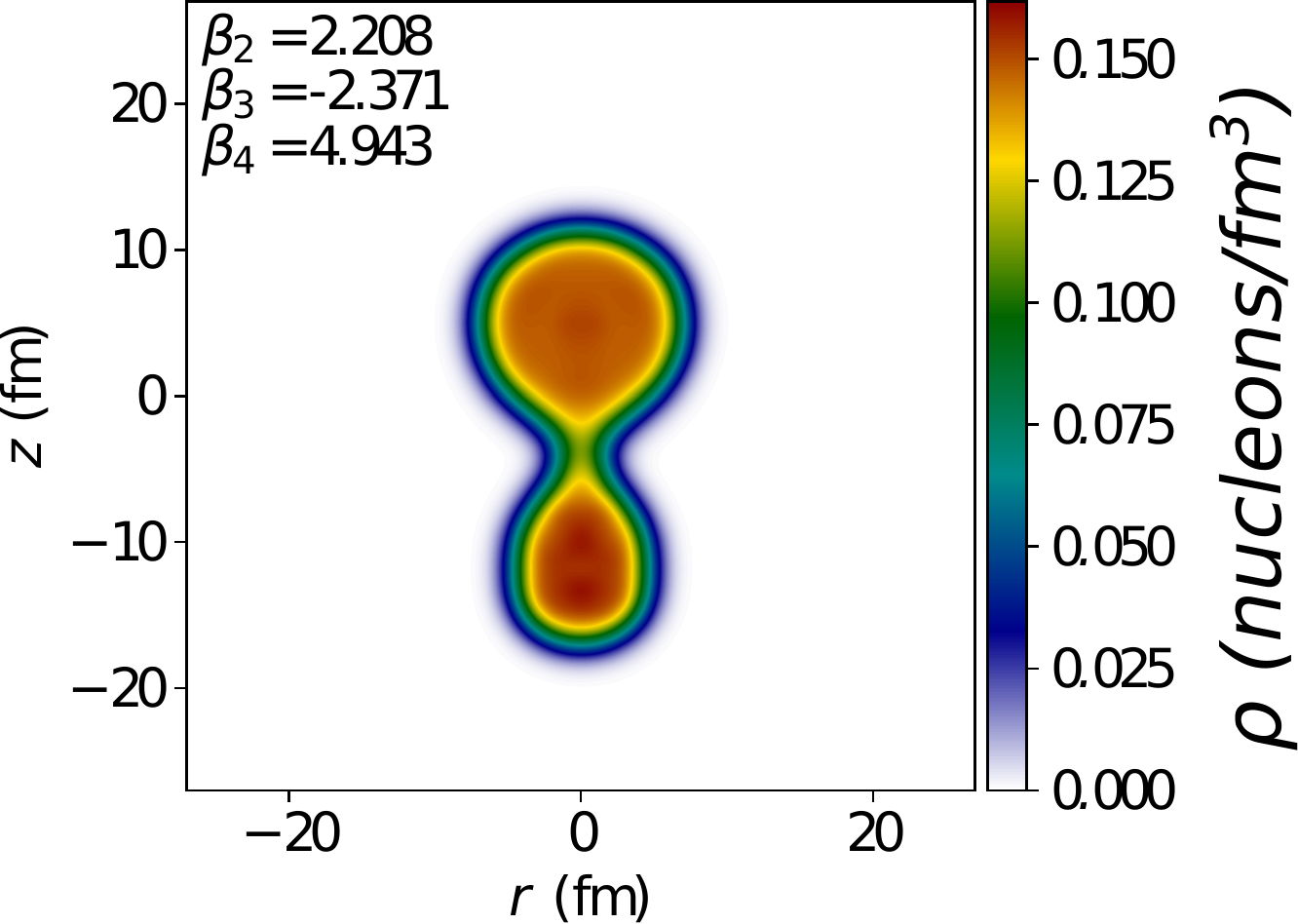}
	\caption{\label{fig:density}Nucleon density  in the $r,z$ plane at the scission point of the asymmetric valley.}
\end{figure}

Figure~\ref{fig:density} shows a nucleon density plot of the scission  point at the end of the asymmetric valley.
 The heavy ($^{208}$Pb) fragment exhibits a small deformation of $\beta_2\simeq0.02$ and $\beta_3=0.1$
 The fact that it is near spherical is attributed to the doubly magic nature of $^{208}$Pb, 
 while its octupole deformation is induced by couplings to its low-lying collective $3^-$ vibrational state. 
  The light ($^{86}$Kr) fragment is more deformed with $\beta_2\simeq0.3$ and $\beta_3\simeq0.12$. 
A relatively shallow and wide symmetric valley (along $Q_{30}=0$) is also observed after the second barrier, with
no clear valley between the ridge which forms the edge of the asymmetric valley and symmetry.

\begin{figure}[h]
	\includegraphics[width=\columnwidth]{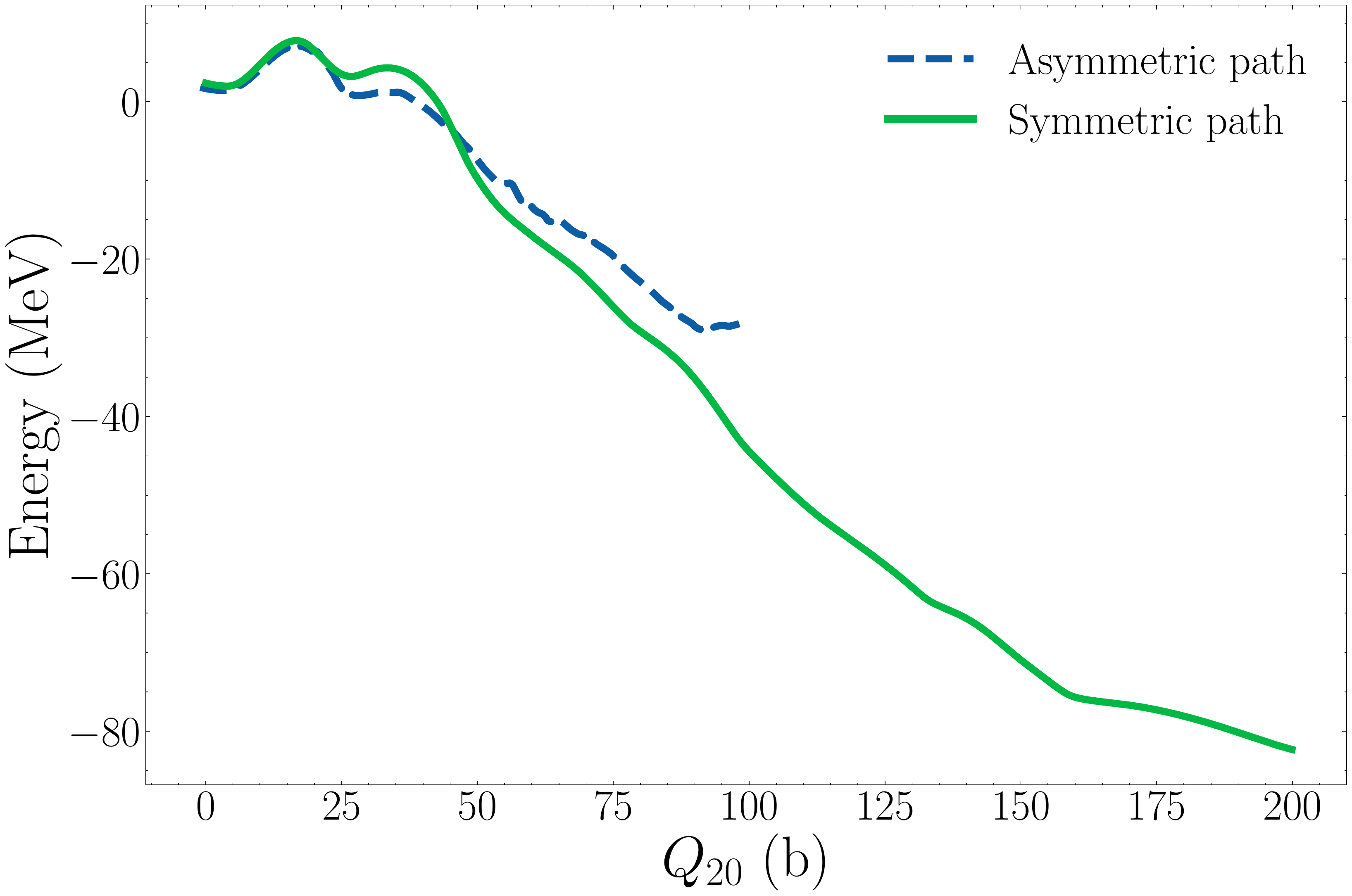}
	\caption{\label{fig_paths}Potential energy along the fission paths. The asymmetric path corresponds to the one plotted in Fig.~\ref{fig_PES} (dashed line) 
	while the symmetric path is that following $Q_{30}=0$.}
\end{figure}

The potential energy as a function of $Q_{20}$ along the asymmetric and symmetric (defined by $Q_{30}=0$) paths is shown in Fig.~\ref{fig_paths}.
All paths exhibit a rapid descent of the potential toward scission due to the large Coulomb repulsion between the fragments. 
The latter is  larger for symmetric fragments, thus leading to a larger slope along the symmetric path.  
Finally, it is also observed that the scission configuration in the asymmetric mode (see also Fig.~\ref{fig:density}) occurs at small $Q_{20}$. 
This can be interpreted as an effect of the heavy fragment ($^{208}$Pb) magicity leading to a compact scission configuration.
Note that, as a result, the production of magic fragments is usually associated with large total kinetic energy (TKE) \cite{hulet1986}.


\section{Quasifission simulations}
\label{sec_QF}
 
\subsection{Initial conditions}
{\setlength{\extrarowheight}{1.5pt}\begin{table}[h]
		\begin{tabular}{cc}
			System &$V_B$ (MeV)\\
			\hline\hline
			$^{48}$Ca$+^{246}$Cf& 235.5 \\
			$^{86}$Kr$+^{208}$Pb&	342.6\\
			$^{126}$Sn$+^{168}$Er&	387.4
		\end{tabular}
		\caption{\label{tab_systems}Systems studied with TDHF in this work. Coulomb barriers $V_B$ are from \cite{swiatecki2005}.  }
\end{table}}
It is unknown prior to running a TDHF calculation if the system will undergo quasifission or another reaction mechanism such as quasielastic scattering or fusion.
It is then necessary to search for quasifission signatures for a range of initial conditions.
These could include various collision partners, centre of mass energies, angular momenta, and orientations in the case of deformed nuclei. 
TDHF investigations of quasifission reactions were performed for the systems given in table \ref{tab_systems}, all leading to the formation of $^{294}$Og compound nucleus in the case of fusion.
Only central collisions were considered to avoid introducing angular momentum into the system, allowing for a comparison (in Sec. \ref{sec_trajs})  with fission paths on the PES that are computed for zero orbital angular momentum. 
Nevertheless, TDHF calculations were performed for a range of energies around the Coulomb barrier $V_B$. 
Note that $^{246}$Cf and $^{168}$Er are prolately deformed. 
Both the tip and side orientations, associated with their deformation axis being parallel and perpendicular to the collision axis, respectively, were considered.

\subsection{Quasifission signatures}

The time-dependent Hartree-Fock (TDHF) approach has been extensively used to investigate quasi-fission mechanisms in the past decade \cite{wakhle2014,oberacker2014,umar2016,umar2015a,hammerton2015,sekizawa2016,guo2018d,zheng2018,godbey2019,simenel2020,yilmaz2020,li2022,wu2022}. 
Three main observables are commonly used to characterise quasifission in such calculations: 
contact times, TKE and final mass asymmetries of the fragments.

\subsubsection{Contact times}

Although contact times are not directly accessible experimentally, they can be inferred through mass-angle distributions \cite{toke1985,durietz2013} as well as neutron emission \cite{hinde1992}.
Quasifission contact times typically exceed few zeptoseconds. 
Quasifission outcomes were then searched for up to 30~zs contact times (1~zs$=1$~zeptosecond$=10^{-21}$~s). 
Longer contact times lead to either fusion or formation of fragments in a process termed ``slow quasifission'' \cite{hinde2021}. 
Earlier TDHF studies of quasifission with actinide targets showed that such long contact times are more easily found with the side orientation \cite{wakhle2014,oberacker2014,umar2016,guo2018d,godbey2019}. 
This can be attributed to the fact that these configurations are more compact at contact, thus favouring  the formation of a compound nucleus. 
Similar results were obtained here in $^{48}$Ca$+^{246}$Cf at above barrier energies and with the side configuration.
Quasi-fission is only found for this system and orientation for a small range of energies  around 210-230 MeV (see Tab.~\ref{CaCfside} in appendix~B).

Figure \ref{fig_contact} shows the evolution of contact times as a function of centre of mass energy normalised to the Coulomb barrier $V_B$. 
The contact time increases rapidly near the energy where contact first occurs. 
For spherical systems, this happens near $V_B$. 
For deformed systems, however, this threshold energy depends on the orientation, with tip collisions reaching contact at lower energies than side collisions.
After a rapid increase at low energy, contact times either plateau or increase slowly with energy. 
The most asymmetric reaction $^{48}$Ca$+^{246}$Cf  leads to the longest contact times ($\sim10-15$~zs for the tip orientation and $\gtrsim15$~zs for the side one). 
As will be discussed later, these longer contact times are associated with more significant mass transfer between the reactants. 

\begin{figure}[h]
	\includegraphics[width=\columnwidth]{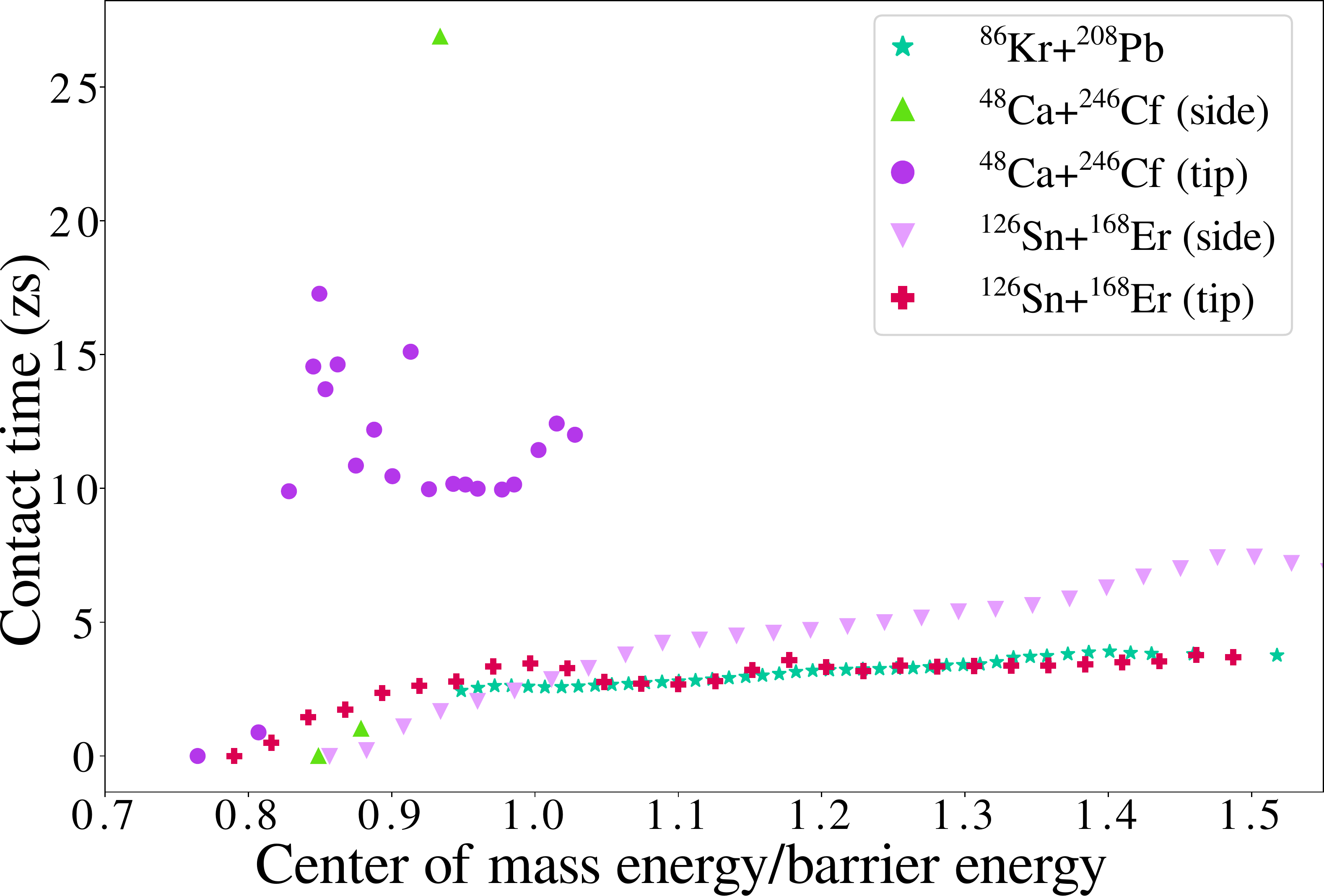}
	\caption{\label{fig_contact}Contact time as a function of centre of mass energy with respect to the Coulomb barrier~$V_B$.}
\end{figure}

\subsubsection{Total kinetic energy}

Another  signature of quasifission is given by the final TKE of the fragments which is expected to follow Viola systematics \cite{viola1985,hinde1987}.
Indeed, initial kinetic energy of the fragments is expected to be fully damped in quasifission.
Nevertheless, fluctuations around Viola systematics could occur, e.g., because of different orientations of deformed nuclei at contact \cite{umar2016,sekizawa2016}.

\begin{figure}[h]
	\includegraphics[width=\columnwidth]{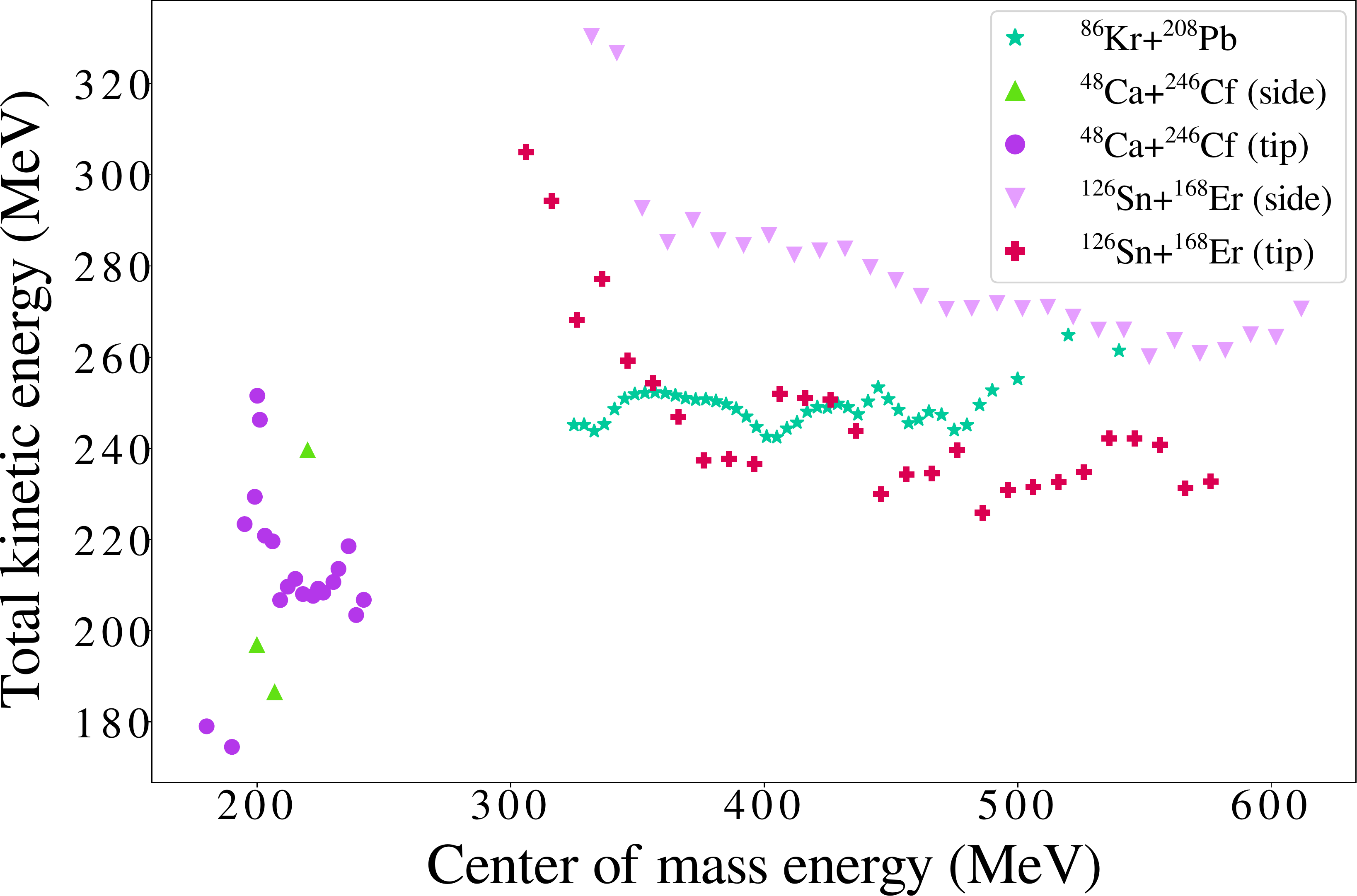}
	\caption{\label{fig_TKE}Total kinetic energy as a function of centre of mass energy.}
	\includegraphics[width=\columnwidth]{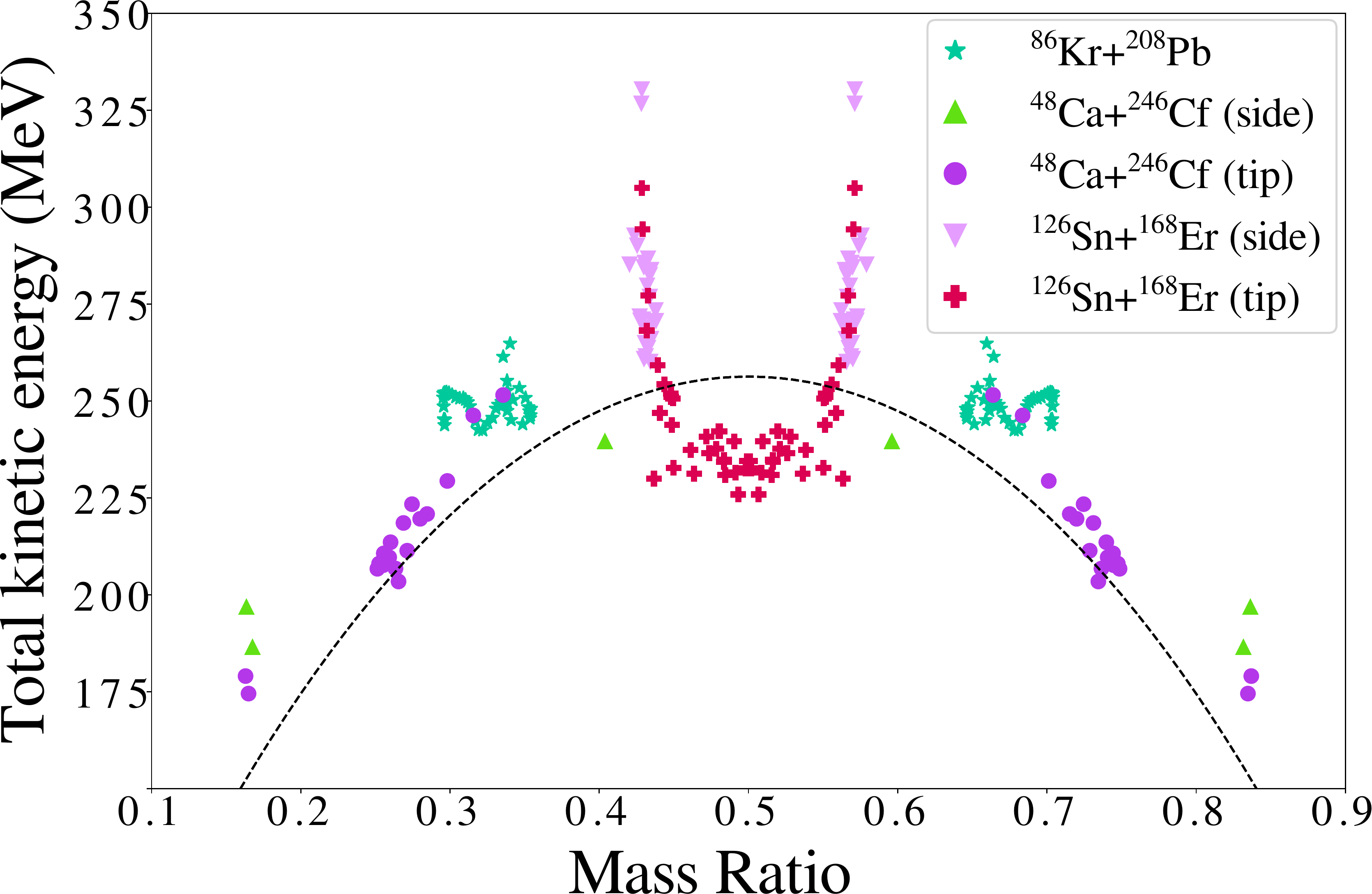}
	\caption{\label{fig_viola}Total kinetic energy as a function of fragment mass ratio (see text). The black dashed line corresponds to  Viola systematics~\cite{hinde1987}. }
\end{figure}

The evolution of the TKE with centre of mass energy is shown in Fig.~\ref{fig_TKE}. 
Each system exhibits essentially no or little dependence of the TKE with energy. 
This indicates that the initial kinetic energy is fully damped. 
This damping is confirmed by their relatively good agreement with the Viola systematics \cite{hinde1987} as seen in Fig.~\ref{fig_viola} showing the TKE as a function of the mass ratio defined as the ratio of  the primary fragment mass over the compound nucleus mass. 

However, a significant deviation from the Viola systematics is  observed in $^{126}$Sn$+^{168}$Er reactions that encounter no or little mass transfer. 
In this case, one expects the influence from the spherical shell effects in the tin region to remain.
The resulting compact configuration at scission is then associated with larger TKE, as  in the case of fission.

\subsubsection{Mass asymmetry}

Finally, quasifission is usually associated with a large mass transfer between the fragments induced by a slow mass equilibration.
Although the production of symmetric fragments requires contact times of the order of $\sim20$~zs \cite{simenel2020}, partial equilibration associated with smaller contact times is often observed. 
The latter can be induced by shell effects in the fragments, preventing further transfer to occur. 
Note that ``inverse quasifission'' mechanisms (leading to more mass asymmetric fragments) induced by shell  \cite{zagrebaev2006} or orientation \cite{kedziora2010} effects, have been also predicted. 

\begin{figure}[h]
	\includegraphics[width=\columnwidth]{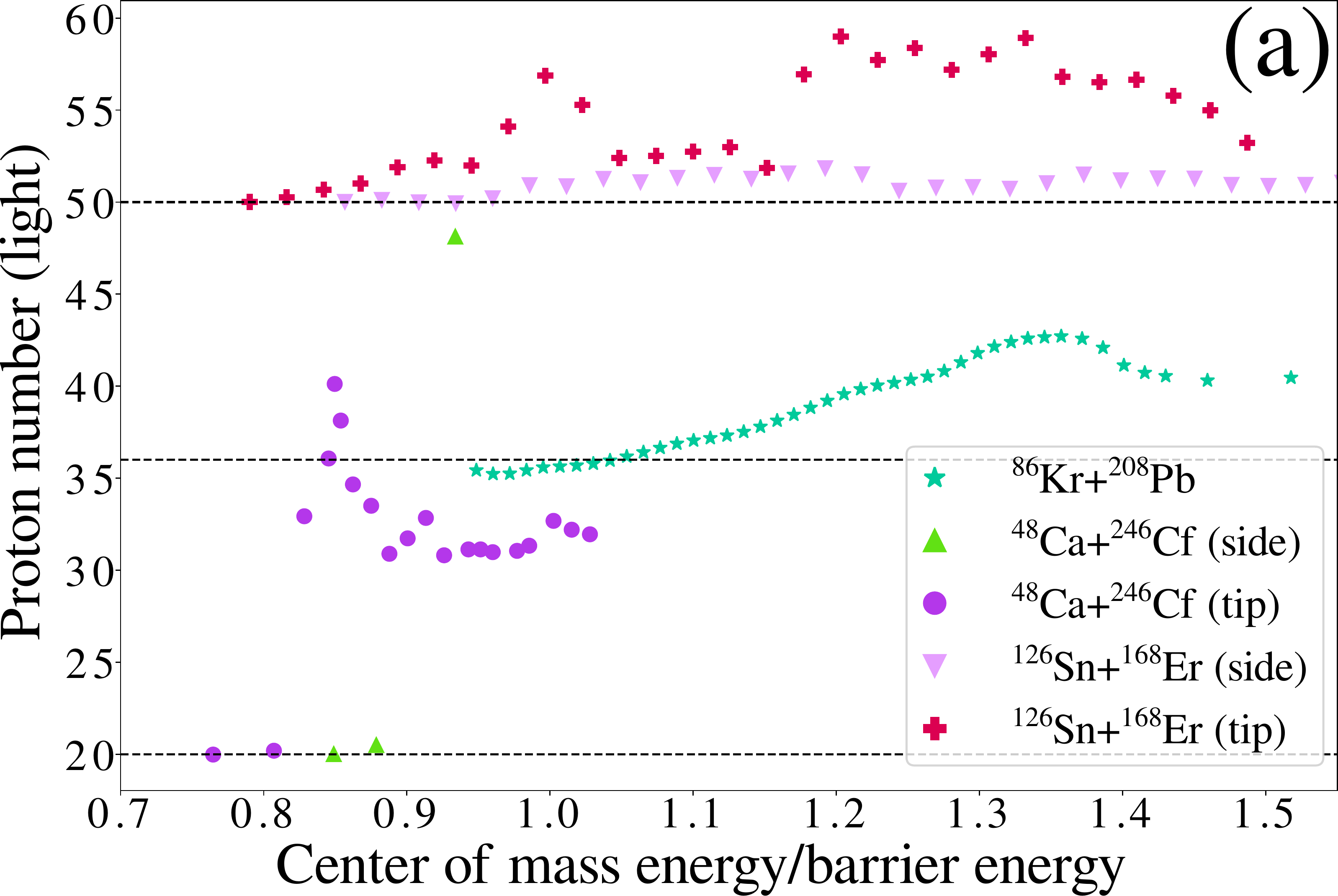}
	\includegraphics[width=\columnwidth]{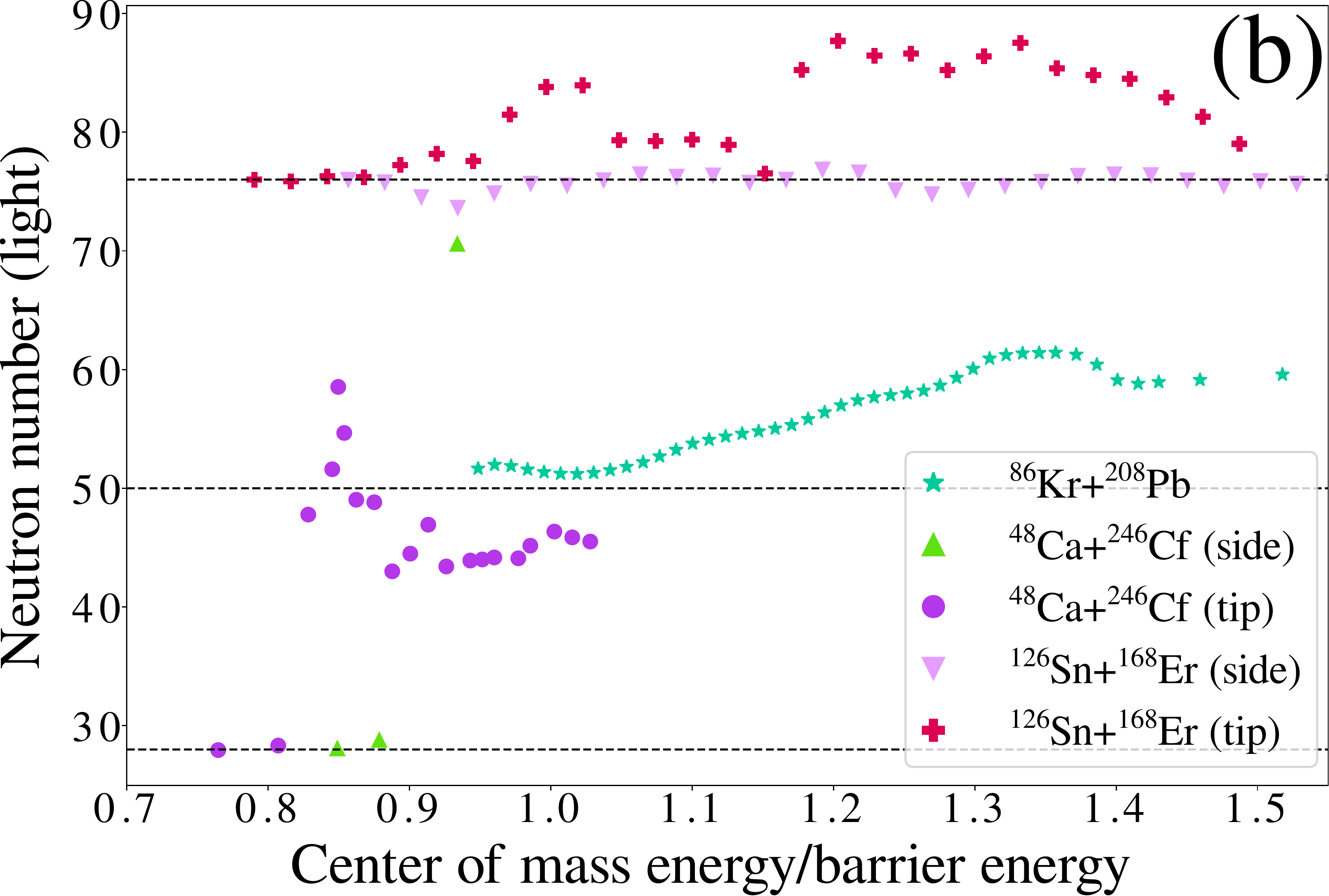}
	\caption{\label{fig_numbers} (a) Proton and (b) neutron numbers of the final (light) primary fragments. Horizontal lines correspond  to the proton and neutron numbers of the original light fragments.}
\end{figure}

The number of protons and neutrons of the outgoing primary fragments produced in quasifission are shown in Fig.~\ref{fig_numbers} as a function of centre of mass energy.
All systems encounter some degree of mass equilibration, although the  $^{126}$Sn$+^{168}$Er side collisions barely lead to any transfer. 
This is interpreted as an influence of the magic shell $Z=50$ in tin iosotopes. 
Collisions with the tip of $^{168}$Er, however, seem to force the Sn fragments out of their magic stability, leading to near symmetric primary fragments. 

$^{48}$Ca$+^{246}$Cf  collisions are those that lead to the largest amount of mass transfer. 
This is compatible with the larger contact times obtained for this system (see Fig.~\ref{fig_contact}). 
The lowest energy quasifissions seem to favour the formation of heavy fragments near the  $^{208}$Pb doubly magic nucleus.
At higher energy, however, the heavy fragment is found typically with $Z_H\sim86$ protons and $N_H\sim131$ neutrons, while the light one has $Z_L\sim31$ and $N_L\sim44$.
Possible shell effects responsible for the formation of these fragments are the octupole deformed shell effects at $Z=84$ and 88 (see, e.g., \cite{butler1996}), and elongated shell effects at $N=42-46$ \cite{macchiavelli1988,nazarewicz1985}.

Finally, a relatively smooth transition is observed in the $^{86}$Kr$+^{208}$Pb system.
At lower energies, little transfer is observed, that can be interpreted as an influence of the doubly magic $^{208}$Pb.
With increasing energies, however, the heavy fragment is formed with $Z_H\sim76$ and $N_H\sim117$, corresponding to a light fragment with $Z_L\sim42$ and $N_L\sim60$. 
This repartition could be influenced by elongated deformed shell effects near $Z=42$ (see, e.g., supplemental material of \cite{scamps2019}). 
In both $^{86}$Kr$+^{208}$Pb and $^{48}$Ca$+^{246}$Cf, the shell effects associated with $^{208}$Pb seem to rapidly wash out with energy. 


\subsection{Quasifission trajectories in $Q_{20}-Q_{30}$ plane}
\label{sec_trajs}
\begin{figure}[h]
	\centering
	\begin{subfigure}[b]{\columnwidth}
		\includegraphics[width=\columnwidth]{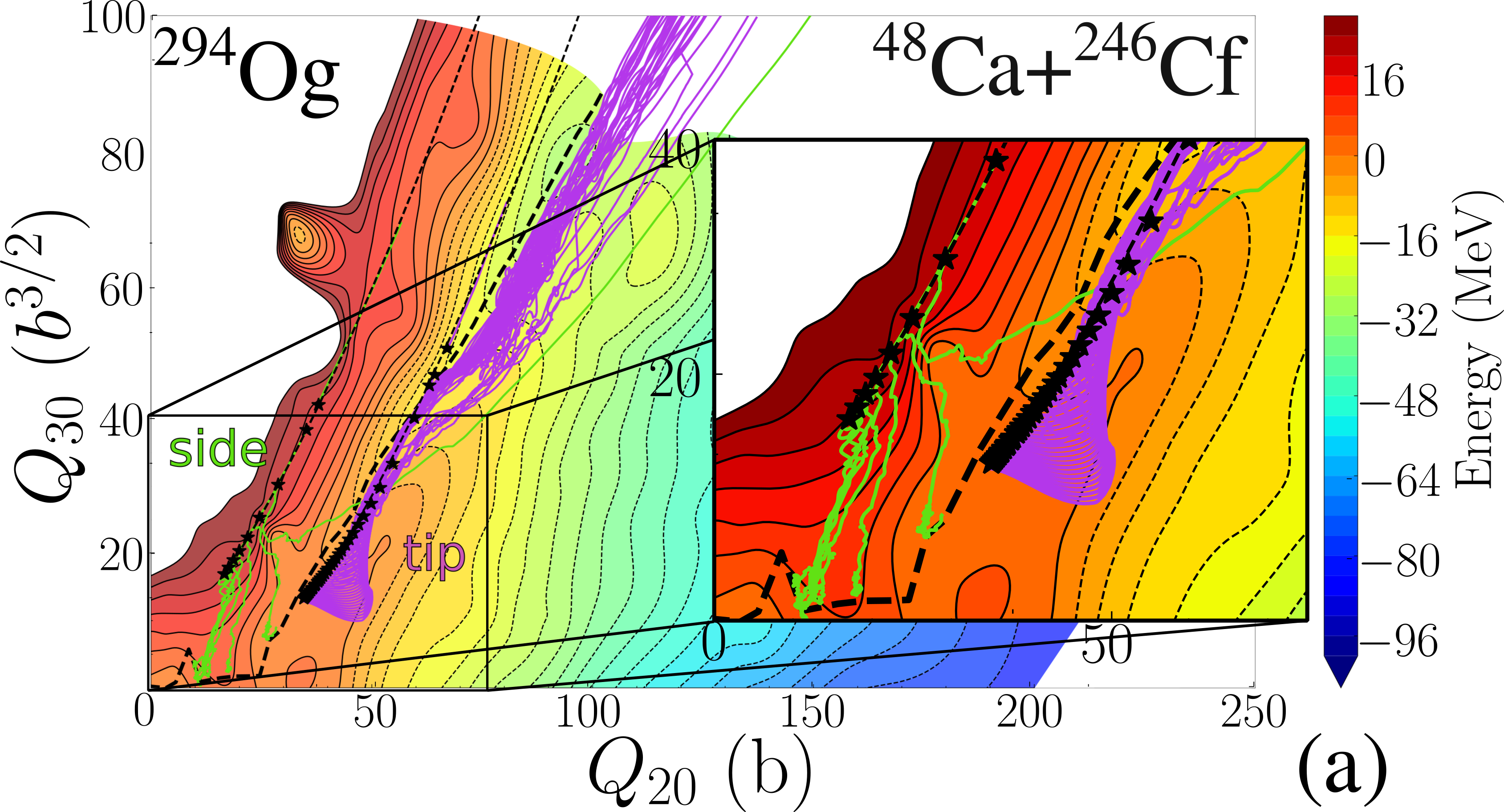}
	\end{subfigure}
	\begin{subfigure}[b]{\columnwidth}
		\includegraphics[width=\columnwidth]{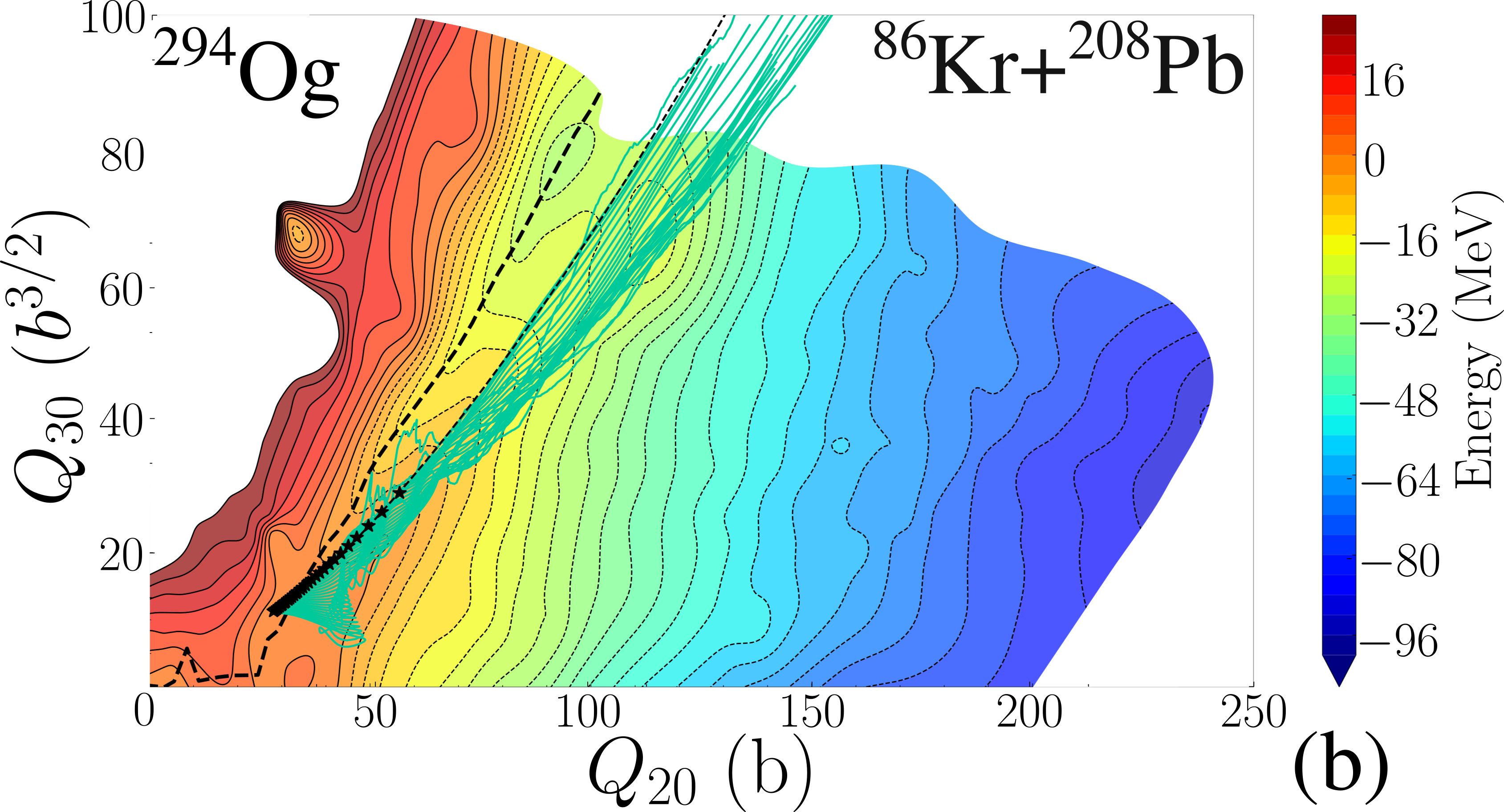}
	\end{subfigure}
	\begin{subfigure}[b]{\columnwidth}
		\includegraphics[width=\columnwidth]{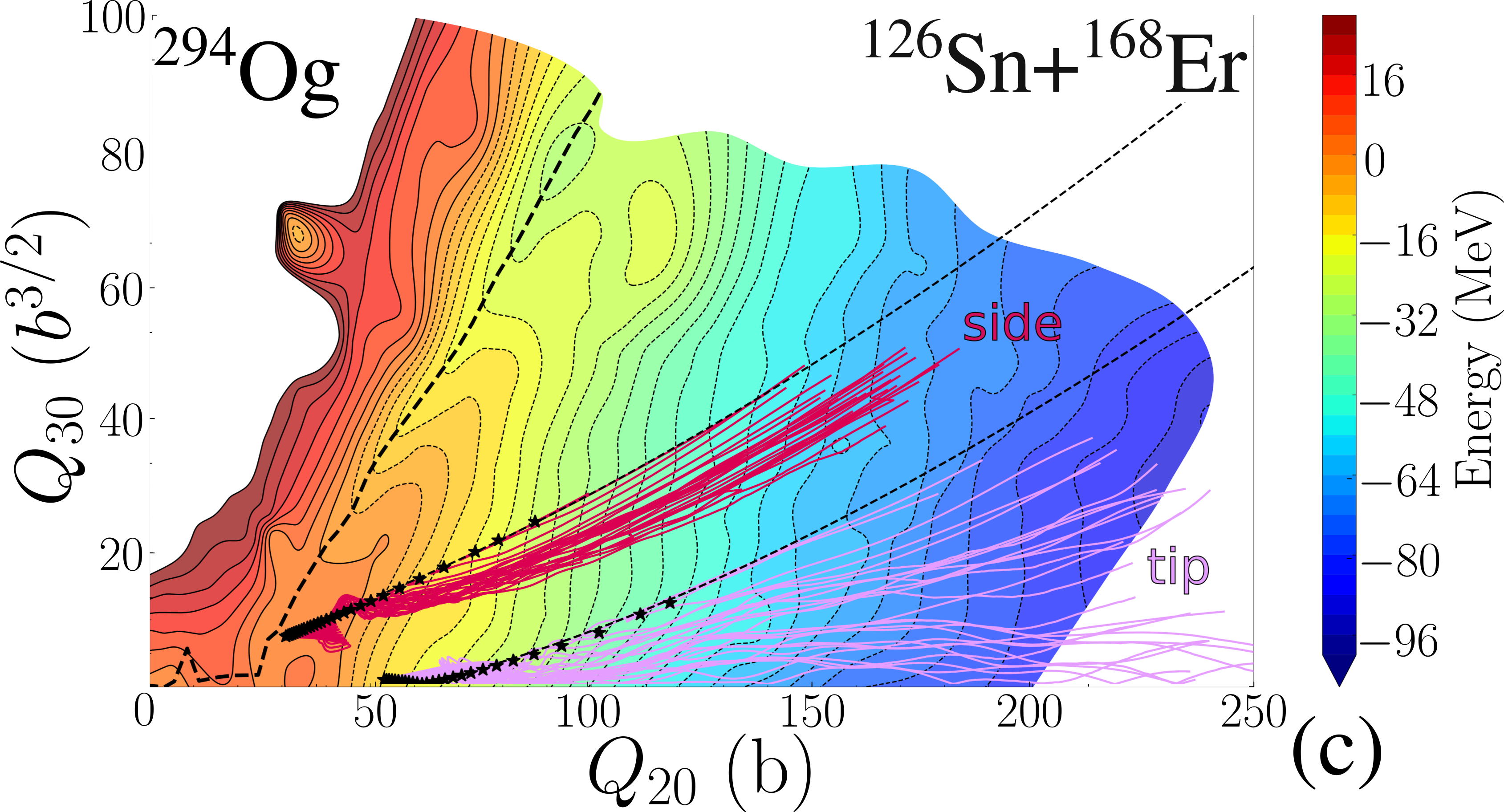}
	\end{subfigure}

	\caption{\label{fig_alltrajs} Overlay of $Q_2-|Q_3|$ trajectories of (a) $^{48}$Ca$+^{246}$Cf, (b)  $^{86}$Kr$+^{208}$Pb, and (c) $^{126}$Sn$+^{168}$Er calculations on the zero-temperature PES of $^{294}$Og. The trajectories are only drawn after the initial kinetic energy is fully dissipated (entry points, represented by stars) until scission. The dotted lines correspond to the incident kinematic trajectories, which are identical for all reactions of the same entrance channel. The dashed line represents the asymmetric fission path. The inset in (a) is a zoom of the compact region of the PES.}
\end{figure}

In the interest of comparison between fission and quasifission modes, the quasifission ``trajectories'' have been determined in the $Q_{20}-|Q_{30}|$ plane and overlaid with the PES of Fig.~\ref{fig_PES}. 
The results are plotted in Fig.~\ref{fig_alltrajs} for each system.
The primary goal of this comparison is to see if these trajectories are affected by the PES topology, keeping in mind that the PES has been determined at zero temperature while finite excitation energies are expected in quasifission. 
In addition, the axial symmetry that was assumed to construct the PES is broken in the TDHF initial conditions for $^{48}$Ca$+^{246}$Cf and $^{126}$Sn$+^{168}$Er with side orientation. 

Each trajectory can be separated into three parts: $(i)$ incoming trajectory determined by kinematics, $(ii)$ fragments in contact, and $(iii)$ post scission outgoing trajectory (also determined by kinematics). 
The incoming trajectory roughly obeys $Q_{30}\propto\sqrt{Q_{20}-Q_{20}^{(frag.)}}$, where $Q_{20}^{(frag.)}$ is the quadrupole moment of the incoming deformed nucleus (see appendix~A) and ends at the ``entry point''.
The latter is obtained when the initial kinetic energy is fully dissipated. 
It also corresponds to the neck density reaching roughly saturation density. 
Here, it is determined from the first time when the neck density exceeds 0.14~fm$^{-3}$. 

\subsubsection{$^{48}$Ca$+^{246}$Cf}

Figure \ref{fig_alltrajs}(a) shows the trajectories for $^{48}$Ca$+^{246}$Cf. 
The incoming trajectories for side and tip orientations are parallel to each other, with the side one associated with smaller $Q_{20}$, and thus leading to more compact configurations. 
The entry points for the side orientation have $Q_{20}$ values smaller than the second barrier (located at $Q_{20}\simeq33$~b and $Q_{30}=0$~b$^{3/2}$), and a potential gradient essentially driving the system towards smaller asymmetries, thus trapping the system into a deformation close to the ground-state one and leading to fusion. 
Quasifission is found for the side orientation in a small energy range between quasi-elastic and fusion. In this case the entry point is located in a region of the PES with a large gradient along the $Q_{20}$ axis [see also inset of Fig.~\ref{fig_alltrajs}(a)], driving the system towards more elongation rather than towards fusion. 

The entry points for the tip orientations, however, are always more asymmetric and elongated than the second barrier and therefore do not lead to fusion. 
Instead, the system remains trapped in the asymmetric valley. Although the $^{48}$Ca$+^{246}$Cf quasifission paths for the tip orientation are found at slightly larger $Q_{20}$
 than the asymmetric fission path (dashed line), this is an indication that the system is affected by similar shell effects in both mechanisms. 

\subsubsection{$^{86}$Kr$+^{208}$Pb}

The $Q_{20}-Q_{30}$ trajectories in the case of $^{86}$Kr$+^{208}$Pb collisions are shown in Fig.~\ref{fig_alltrajs}(b).
The incoming trajectory essentially follows the ridge between the asymmetric valley and more symmetric modes.
As a result, the entry points are also located along that ridge. 
As in the $^{48}$Ca$+^{246}$Cf tip orientation case, most of these entry points are located at larger elongations than the second barrier, and therefore are not able to find a path to fusion. 
Only the highest centre of mass energies lead to entry points slightly more compact than the second barrier. 
However, at these energies the shell effects responsible to the structures in the PES have potentially disappeared.

Finally, despite apparent initial fluctuations near the entry points that lead to some exploration of the asymmetric valley, all outgoing trajectories end up further away from the asymmetric mode than the entrance channel.
As a result, there is no clear correlation between the PES and the quasifission trajectories in this system. 
Once again, this could be due to the larger excitation energies induced by the large initial kinetic energy necessary to overcome the stronger Coulomb barrier in more symmetric systems. 

\subsubsection{$^{126}$Sn$+^{168}$Er}

Finally, quasifission trajectories are shown for  $^{126}$Sn$+^{168}$Er tip and side orientations in Fig.~\ref{fig_alltrajs}(c).
The incoming trajectories explore configurations away from the symmetric and asymmetric modes. 
The tip orientation is associated with smaller values of the octupole moment than the side orientation, and even leads to entry points with $Q_{30}\simeq0$ at highest energies. 
Neither orientation, however, reaches more compact configurations than the second barrier. 
All collisions lead to quasifission with smaller $Q_{30}$ in the exit channel. 
Tip orientations are even able to produce near-symmetric fragments. 

Overall, no clear correlation between the PES and quasifission trajectories is observed. 
This could be due to the fact that the PES is relatively flat in the $Q_{30}$ direction in the region explored by $^{126}$Sn$+^{168}$Er reactions.  
In addition, being the most symmetric reaction, it is also the one with the highest Coulomb barrier. 
As a result, the amount of excitation energy deposited into the system in the entry point is very large 
and could wash out shell effects.

\section{Discussion}

We see in Fig. \ref{fig_alltrajs}  that the three reactions  cover much of the $Q_{20}-Q_{30}$ plane that is relevant for the PES.
Naturally, this does not mean that all reactions are sensitive to the structure of the PES itself. 
The more symmetric collisions require more energy to overcome the Coulomb barrier.
Thus more energy is converted into excitation energy that may hinder, if not kill shell effects and thus structures on the PES. 
Nevertheless, the most asymmetric reactions seem to be sensitive to these structures, such as the second barrier, direction of the gradient of the PES near the entry point, as well as the asymmetric valley. 

We also observe that the trajectories for different entrance channels (defined as colliding nuclei with a specific orientation) essentially do not overlap with each other, despite the broad range of energies considered for each system. 
This indicates that the reaction mechanisms (in terms of shape evolution) are more sensitive to the initial configuration than to the energy of the collision. 

\begin{figure}
	\includegraphics[width=\columnwidth]{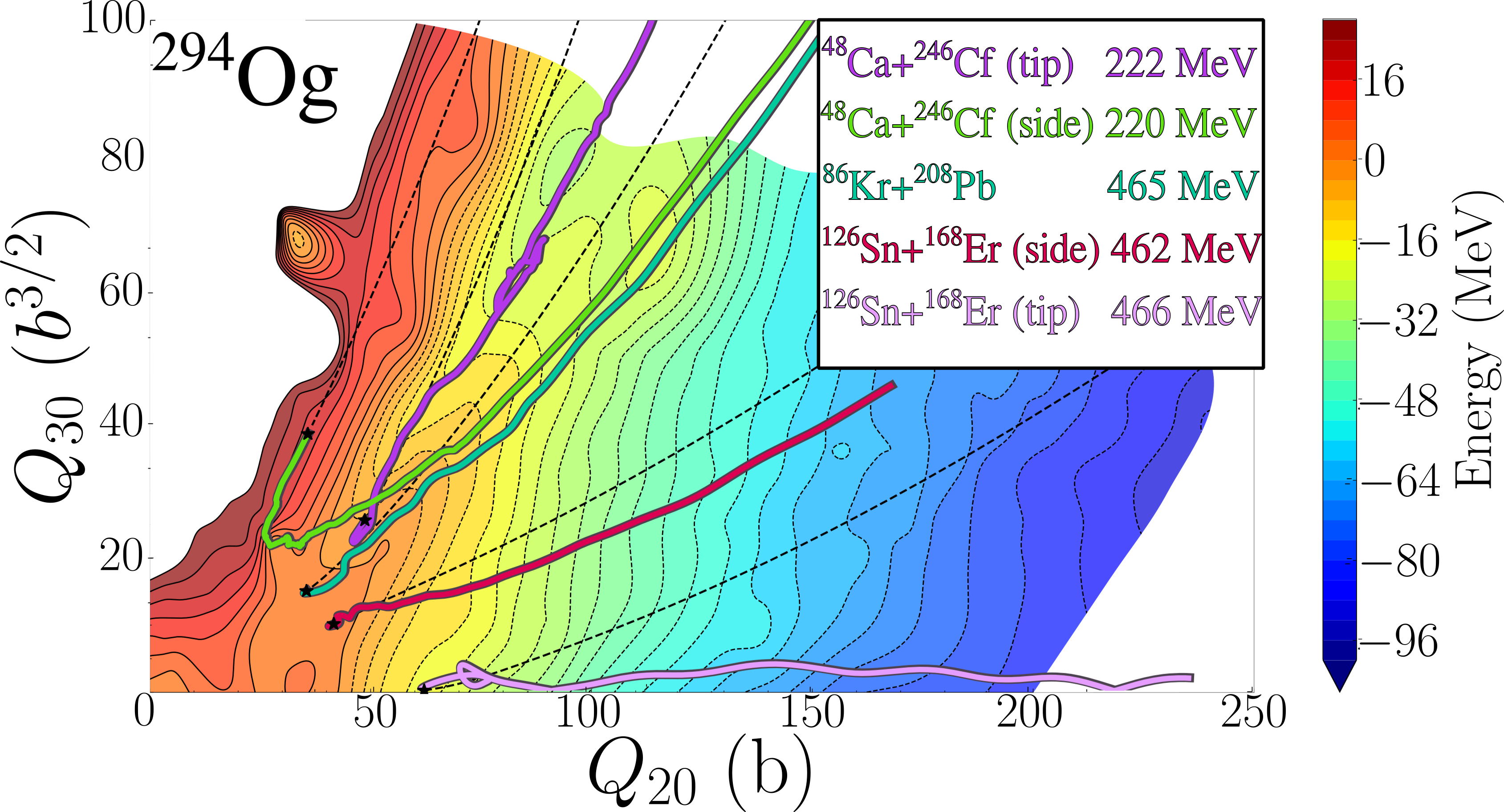}
	\caption{\label{fig_selectrajs}  Overlay of one quasifission trajectory for each entrance channel on the zero-temperature PES of $^{294}$Og. Dotted black lines represent entrance trajectories, stars indicate the entrance point on the PES of each trajectory.
}
\end{figure}

\begin{figure}
	\includegraphics[width=\columnwidth]{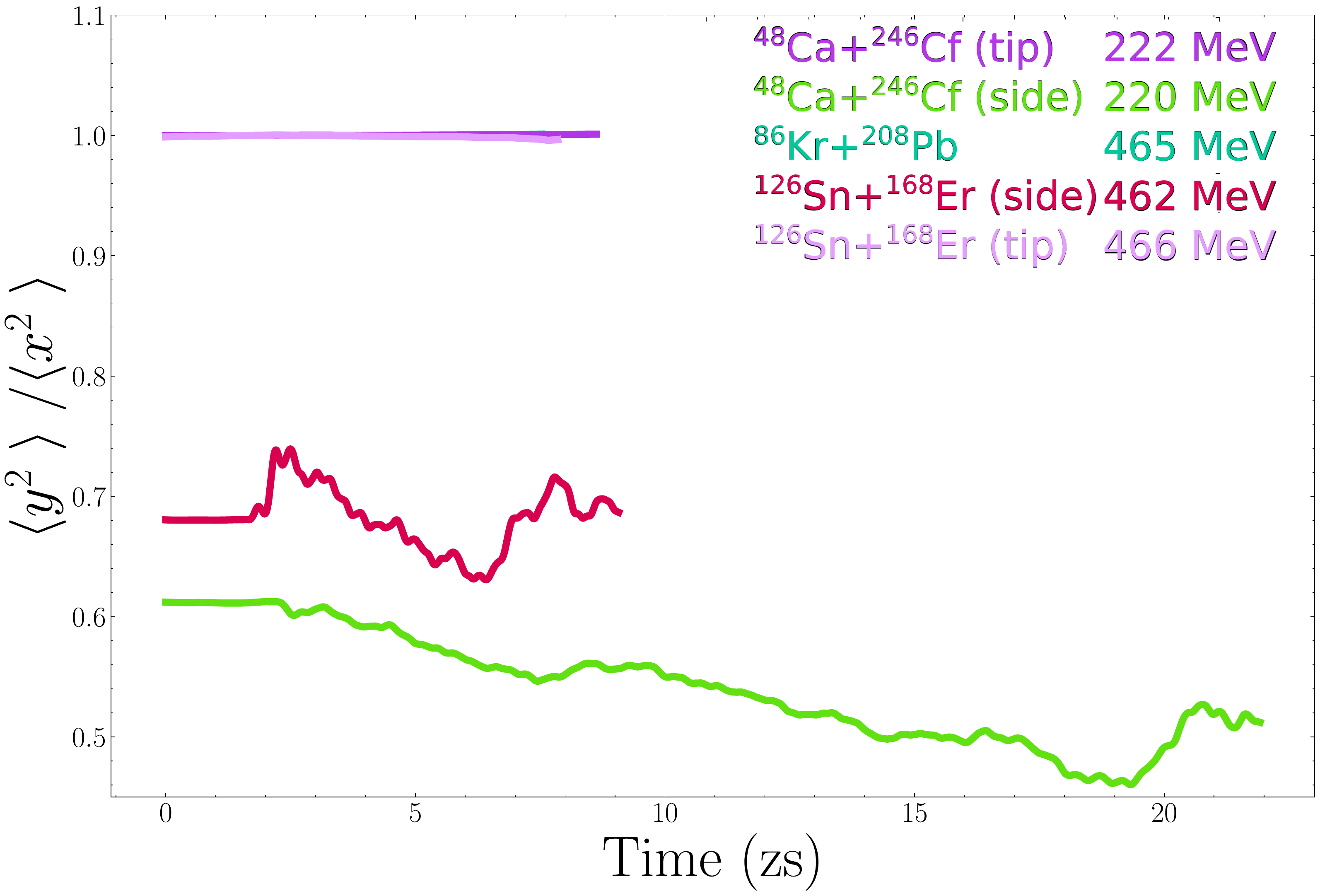}
	\caption{\label{fig_axial}  Evolution of $\langle y^2\rangle/\langle x^2\rangle$ as a function of time for the systems considered in Fig.~\ref{fig_selectrajs}. Systems with an axial symmetry are expected to give a value of 1.}
\end{figure}

Selected trajectories for each system and orientation are shown in Fig. \ref{fig_selectrajs}.
The $^{48}$Ca$+^{246}$Cf side orientation at $E_{c.m.}\simeq220$~MeV leads to the longest contact time ($\sim27$~zs) found in the entire set of calculations. 
Soon after the entry point the system follows a steep descent in the PES before going towards scission [see also inset of Fig.~\ref{fig_alltrajs}(a)]. 
Here, the path to scission is slightly more symmetric than the asymmetric path. 
This could  indicate that the PES is not entirely relevant for this outgoing trajectory as the latter assumes axial symmetry while the side orientation breaks this symmetry.
Indeed, the evolution of $\langle y^2\rangle/\langle x^2\rangle$ shown in Fig.~\ref{fig_axial} indicates that collisions with side orientations remain non-axial during the entire reaction. (A value different than 1 is a signature for a non-axial shape for collisions along the $z-$axis.) 
Moreover, dynamical effects may take the system away from the adiabatic asymmetric path. 
The $^{48}$Ca$+^{246}$Cf tip orientation (which is axially symmetric for central collisions) is shown at the same energy  in Fig. \ref{fig_selectrajs} for comparison. In this case, the outgoing trajectory essentially follows the asymmetric valley. 

The $^{86}$Kr$+^{208}$Pb trajectory is shown for $E_{c.m.}\simeq465$~MeV. 
It corresponds to the largest mass transfer for this reaction ($\sim6.7$ protons and $\sim11.4$ neutrons from the heavy fragment to the light one). 

Figure~\ref{fig_selectrajs} also shows the trajectories for the side and tip collisions of $^{126}$Sn$+^{168}$Er at $E_{c.m.}\simeq462$ and 466~MeV, respectively. 
These energies also correspond to the largest charge transfer for this system: 1.8 (9.0) protons for the side (tip) orientation. 
Although the largest charge transfer is obtained at similar energies, the amount of transfer is very different between both orientations.
In particular, spherical shell effects in the magic Sn ($Z=50$) collision partner seem to hinder transfer for side orientation at all energy. 
For tip collisions, however, the entry point reaches $Q_{30}\sim0$ and, as a result, full mass and charge equilibration is observed in the exit channel. 

The quasifission trajectories in Figs.~\ref{fig_alltrajs} and \ref{fig_selectrajs} are plotted up to scission, defined as the neck density becoming smaller than $\rho_0/2=0.08$~fm$^{-3}$. 
We see that scission in $^{48}$Ca$+^{246}$Cf and $^{86}$Kr$+^{208}$Pb quasifission  reactions occurs beyond the scission line of the PES.
This does not necessarily indicate a difference between fission and quasifission mechanisms as the PES is obtained with an adiabatic approximation that breaks down near scission due to non-adiabatic dynamical effects \cite{simenel2014a}.  Non-adiabatic effects are expected to affect the scission configuration in fission and quasifission in a similar way. 
Comparing scission configurations in fission and quasifission would require full time-dependent calculations of fission dynamics \cite{simenel2014a,goddard2015,scamps2015a,bulgac2016,ren2022} that are beyond the scope of this work. 
Nevertheless, we note that, in the case of $^{126}$Sn$+^{168}$Er, most quasifission trajectories scission before the scission line, except for the most symmetric exit channels. 
This can be interpreted as an effect of spherical magic shell at $Z=50$ in the light fragment, leading to more compact configurations at scission. This is in agreement with the observation of higher TKE for these reactions in Fig.~\ref{fig_TKE}.

\section{Conclusion}

\label{sec_conc}

Quasifission mechanisms have been studied with the TDHF approach in several reactions forming $^{294}$Og as a compound nucleus.
The exit channel strongly depends on the mass asymmetry of the collision partners, as well as their initial orientation.
However, it exhibits only a small dependence with the energy of the reaction. 

Trajectories in the $Q_{20}-Q_{30}$ plane were extracted and compared with the PES computed for axial shapes.
Properties of the most asymmetric collision ($^{48}$Ca$+^{246}$Cf) could be interpreted in terms of features of the PES. 
In particular, the topology near the entry point determines whether the system fuses or encounters quasifission along the asymmetric valley. 

Spherical shell effects associated with the $Z=50$ magic gap hinder charge equilibration in the most symmetric reaction, $^{126}$Sn$+^{168}$Er, resulting in larger TKE and more compact scission configurations. 
Nevertheless, collisions with the tip of $^{168}$Er are able to take the tin collision partner away from magicity, leading to the production of near symmetric fragments. 

More symmetric collisions require more energy to reach contact. 
The resulting excitation energy could wash out shell effects responsible for the structure of the PES. 
For these reactions, the zero-temperature PES may not be the best predictor of energy variation with shape changes.
Instead a PES accounting for excitation energy could be used. 
Calculations of free energy variation with shape at finite temperature (see, e.g., \cite{mcdonnell2014,pei2009a,zhao2019b,li2023}) show that new valleys emerge and modes shift to different positions. 
The dissipation process outlined in \cite{zhao2022a,zhao2022b} on the way to scission is also expected to affect these trajectories.
The development of PES at finite excitation energy and finite angular momentum for quasifission studies will be the purpose of future works. 

\begin{acknowledgments}
The authors would like to acknowledge the valuable contributions of R\'emi Bernard, especially on the calculations of potential energy surfaces. 
Contribution from Daniel Lee on the discussions of trajectories is acknowledged. 
This work has been supported by the Australian Research Council Discovery Project (project number DP190100256).
Computational resources were provided by the Australian Government through the National Computational Infrastructure (NCI) under the ANU Merit Allocation Scheme, the Adapter scheme, and the National Committee Merit Allocation Scheme. P. McGlynn acknowledges the support of the Australian National University through the Deakin PhD scholarship and the Dean's Merit Scholarship in science.
\end{acknowledgments}

\appendix
\section*{Appendix A: Incoming trajectory in the $Q_{20}-Q_{30}$ plane}
At large distances, in order to approximate the trajectory of the incoming nuclei, their $Q_{20}$ and $Q_{30}$ moments can be approximated by treating the nuclei as point masses $m_1$ and $m_2$ located at $z_1$ and $z_2$ from the center of mass respectively. Then
\begin{align}
	Q_2&=2\sqrt{\frac{5}{16\pi}}m_2z_2^2\left(1+\frac{m_2}{m_1}\right)\\
	Q_3&=2\sqrt{\frac{7}{16\pi}}m_2z_2^3\left(1-\frac{m_2^2}{m_1^2}\right)\\
	&=\frac{\sqrt{14}\,\pi^{1/4}}{5^{3/4}}\frac{\alpha-1}{\sqrt{m_T\alpha}}Q_2^{3/2},
\end{align}
where $\alpha=m_1/m_2,m_t=m_1+m_2$. This gives the general dependence of the incoming trajectory. 
There is an offset however, associated with non-spherical nuclei, which can be accounted for to give the following more complete approximation:
\begin{align}
	Q_3=&\frac{\sqrt{14}\pi^{1/4}}{5^{3/4}}\frac{\alpha-1}{\sqrt{\alpha m_T}}\left(Q_2+Q_{eff}\right)^{3/2}\\
	&+\frac{\sqrt{14}\pi^{1/4}}{5^{3/4}}\frac{3\alpha}{\sqrt{\alpha m_T}}\left(Q_2+Q_{eff}\right)^{1/2}Q_{eff}
\end{align}
where $Q_{eff}=0$ when both nuclei are spherical, $Q_{eff}=Q_2^{frag.}$ in a tip collision, and $Q_{eff}=-Q_2^{frag.}/2$ in a side collision, 
with $Q^{frag.}$ the groundstate quadrupole moment of the deformed reaction partner.

\section*{Appendix B: Details of TDHF results}

The following tables provide details of the TDHF results used to produce the figures of the manuscript.
The data are for primary fragments, i.e., prior to subsequent decay.
Data are  presented for calculations leading to two fragments in the exit channel, unless they are labelled as ``fusion'' which in this context means contact times exceeding 30~zs.

\begin{table}[ht]
\begin{ruledtabular}
\begin{tabular}{ccccccc}
$E_{cm}$ &Contact& $Z_H$& $N_H$&$Z_L$&$N_L$&TKE \\
\hline
199.92 & 0.0 & 97.99 & 147.89 & 20.01 & 28.08 & 196.86 \\ 
206.92 & 1.021 & 97.48 & 147.03 & 20.5 & 28.79 & 186.5 \\ 
209.92 & 2.172 & 96.42 & 146.17 & 21.57 & 29.62 & 185.6 \\ 
214.92 & 14.10 & 81.96 & 123.98 & 36.03 & 51.87 & 239.21 \\ 
219.92 & 26.88 & 69.86 & 105.4 & 48.13 & 70.59 & 239.58 \\ 
233.92 & fusion &   &  &  &  &   \\ 
\end{tabular}
\caption{\label{CaCfside}Fragments produced in $^{48}$Ca$+^{246}$Cf with the side orientation. Energies ($E_{cm}$ and TKE) are in MeV and contact times (Contact) in zeptoseconds.
The subscripts $H$ and $L$ stand for heavy and light fragments, respectively. 
}
\end{ruledtabular}
\end{table}

\begin{table}[ht]
\begin{ruledtabular}
\begin{tabular}{ccccccc}
$E_{cm}$ &Contact& $Z_H$& $N_H$&$Z_L$&$N_L$&TKE \\
\hline
180.08 & 0.0 & 98.02 & 148.04 & 19.98 & 27.94 & 179.01 \\ 
190.08 & 0.883 & 97.8 & 147.6 & 20.19 & 28.32 & 174.49 \\ 
195.08 & 9.892 & 85.06 & 127.97 & 32.93 & 47.78 & 223.37 \\ 
199.08 & 14.55 & 81.92 & 124.24 & 36.07 & 51.6 & 229.36 \\ 
200.08 & 17.27 & 77.87 & 117.31 & 40.12 & 58.54 & 251.53 \\ 
201.08 & 13.70 & 79.86 & 121.15 & 38.13 & 54.66 & 246.26 \\ 
203.08 & 14.63 & 83.34 & 126.96 & 34.66 & 49.03 & 220.83 \\ 
206.08 & 10.85 & 84.5 & 127.14 & 33.5 & 48.82 & 219.59 \\ 
209.08 & 12.19 & 87.11 & 133.0 & 30.89 & 43.0 & 206.7 \\ 
212.08 & 10.45 & 86.27 & 131.5 & 31.73 & 44.49 & 209.65 \\ 
215.08 & 15.10 & 85.16 & 129.07 & 32.84 & 46.93 & 211.36 \\ 
218.08 & 9.968 & 87.19 & 132.58 & 30.81 & 43.41 & 208.02 \\ 
222.08 & 10.16 & 86.87 & 132.08 & 31.13 & 43.91 & 207.63 \\ 
224.08 & 10.14 & 86.87 & 131.99 & 31.13 & 44.0 & 209.21 \\ 
226.08 & 9.988 & 87.02 & 131.83 & 30.98 & 44.17 & 208.34 \\ 
230.08 & 9.956 & 86.94 & 131.89 & 31.05 & 44.1 & 210.67 \\ 
232.08 & 10.14 & 86.67 & 130.83 & 31.33 & 45.16 & 213.55 \\ 
236.08 & 11.43 & 85.32 & 129.64 & 32.68 & 46.35 & 218.52 \\ 
239.07 & 12.42 & 85.8 & 130.13 & 32.2 & 45.86 & 203.42 \\ 
242.07 & 12.00 & 86.05 & 130.49 & 31.95 & 45.51 & 206.76 \\ 
\end{tabular}
\caption{\label{CaCftip} Same as Tab.~\ref{CaCfside} for $^{48}$Ca$+^{246}$Cf with the tip orientation. }
\end{ruledtabular}
\end{table}

\begin{table}[ht]
\begin{ruledtabular}
\begin{tabular}{ccccccc}
$E_{cm}$ &Contact& $Z_H$& $N_H$&$Z_L$&$N_L$&TKE \\
\hline
324.97 & 2.439 & 82.55 & 124.17 & 35.43 & 51.66 & 245.12 \\ 
328.97 & 2.547 & 82.75 & 123.84 & 35.23 & 51.98 & 245.16 \\ 
332.97 & 2.619 & 82.72 & 123.98 & 35.25 & 51.88 & 243.77 \\ 
336.97 & 2.622 & 82.54 & 124.29 & 35.43 & 51.58 & 245.3 \\ 
340.97 & 2.593 & 82.39 & 124.51 & 35.59 & 51.36 & 248.61 \\ 
344.97 & 2.573 & 82.34 & 124.66 & 35.64 & 51.23 & 250.9 \\ 
348.97 & 2.577 & 82.3 & 124.7 & 35.69 & 51.21 & 251.89 \\ 
352.97 & 2.603 & 82.2 & 124.62 & 35.79 & 51.3 & 252.18 \\ 
356.97 & 2.635 & 82.03 & 124.44 & 35.96 & 51.53 & 252.3 \\ 
360.97 & 2.665 & 81.81 & 124.15 & 36.18 & 51.81 & 252.1 \\ 
364.97 & 2.695 & 81.58 & 123.78 & 36.41 & 52.2 & 251.61 \\ 
368.97 & 2.731 & 81.35 & 123.3 & 36.65 & 52.69 & 251.0 \\ 
372.97 & 2.763 & 81.12 & 122.76 & 36.88 & 53.24 & 250.68 \\ 
376.97 & 2.791 & 80.95 & 122.23 & 37.05 & 53.77 & 250.77 \\ 
380.97 & 2.819 & 80.82 & 121.9 & 37.18 & 54.09 & 250.35 \\ 
384.97 & 2.861 & 80.68 & 121.62 & 37.32 & 54.37 & 249.69 \\ 
388.97 & 2.915 & 80.48 & 121.4 & 37.52 & 54.6 & 248.65 \\ 
392.97 & 2.963 & 80.2 & 121.2 & 37.8 & 54.8 & 246.98 \\ 
396.97 & 3.008 & 79.87 & 120.97 & 38.13 & 55.03 & 244.69 \\ 
400.97 & 3.066 & 79.55 & 120.67 & 38.45 & 55.33 & 242.56 \\ 
404.97 & 3.144 & 79.17 & 120.17 & 38.83 & 55.83 & 242.44 \\ 
408.97 & 3.190 & 78.78 & 119.6 & 39.21 & 56.4 & 244.35 \\ 
412.96 & 3.211 & 78.43 & 119.0 & 39.57 & 56.99 & 245.66 \\ 
416.96 & 3.226 & 78.16 & 118.58 & 39.84 & 57.41 & 248.04 \\ 
420.96 & 3.240 & 77.96 & 118.32 & 40.04 & 57.67 & 249.03 \\ 
424.96 & 3.253 & 77.82 & 118.15 & 40.18 & 57.85 & 248.92 \\ 
428.96 & 3.268 & 77.65 & 118.0 & 40.35 & 58.0 & 249.79 \\ 
432.96 & 3.286 & 77.48 & 117.77 & 40.52 & 58.22 & 249.02 \\ 
436.96 & 3.322 & 77.18 & 117.33 & 40.82 & 58.66 & 247.46 \\ 
440.96 & 3.385 & 76.69 & 116.69 & 41.3 & 59.31 & 250.28 \\ 
444.96 & 3.403 & 76.2 & 115.92 & 41.8 & 60.07 & 253.35 \\ 
448.96 & 3.427 & 75.85 & 115.06 & 42.15 & 60.93 & 250.8 \\ 
452.96 & 3.513 & 75.6 & 114.76 & 42.4 & 61.23 & 248.41 \\ 
456.96 & 3.674 & 75.41 & 114.63 & 42.59 & 61.37 & 245.53 \\ 
460.96 & 3.711 & 75.34 & 114.6 & 42.66 & 61.4 & 246.32 \\ 
464.96 & 3.736 & 75.29 & 114.57 & 42.71 & 61.43 & 248.0 \\ 
469.96 & 3.809 & 75.42 & 114.73 & 42.58 & 61.26 & 247.35 \\ 
474.96 & 3.872 & 75.9 & 115.57 & 42.09 & 60.42 & 244.0 \\ 
479.96 & 3.907 & 76.87 & 116.88 & 41.13 & 59.11 & 245.1 \\ 
484.96 & 3.850 & 77.27 & 117.19 & 40.73 & 58.81 & 249.53 \\ 
489.96 & 3.821 & 77.45 & 117.04 & 40.55 & 58.95 & 252.7 \\ 
499.96 & 3.807 & 77.69 & 116.86 & 40.31 & 59.13 & 255.22 \\ 
519.95 & 3.760 & 77.53 & 116.41 & 40.46 & 59.58 & 264.82 \\ 
539.95 & 3.699 & 77.91 & 117.4 & 40.08 & 58.6 & 261.42 \\ 
559.95 & 3.600 & 77.98 & 118.06 & 40.02 & 57.94 & 253.03 \\ 
569.95 & 3.553 & 78.57 & 118.58 & 39.43 & 57.42 & 244.83 \\ 
579.95 & 3.476 & 79.5 & 119.21 & 38.51 & 56.79 & 243.14 \\ 
584.95 & 3.507 & 79.98 & 119.65 & 38.02 & 56.35 & 236.79 \\ 
589.95 & 3.594 & 79.42 & 119.24 & 38.57 & 56.75 & 240.78 \\ 
599.95 & 3.685 & 77.57 & 118.09 & 40.43 & 57.9 & 252.35 \\ 
609.94 & 4.403 & 76.68 & 115.13 & 41.32 & 60.87 & 246.15 \\ 
619.94 & 5.266 & 75.86 & 115.26 & 42.14 & 60.73 & 245.43 \\ 
629.94 & 5.982 & 80.22 & 119.49 & 37.78 & 56.51 & 228.06 \\ 
\end{tabular}
\caption{\label{KrPb} Same as Tab.~\ref{CaCfside} for $^{86}$Kr$+^{208}$Pb.}
\end{ruledtabular}
\end{table}

\begin{table}[ht]
\begin{ruledtabular}
\begin{tabular}{ccccccc}
$E_{cm}$ &Contact& $Z_H$& $N_H$&$Z_L$&$N_L$&TKE \\
\hline
331.89 & 0.0 & 67.98 & 99.97 & 50.01 & 75.99 & 330.43 \\ 
341.89 & 0.216 & 67.87 & 100.07 & 50.12 & 75.78 & 326.79 \\ 
351.89 & 1.106 & 67.99 & 101.3 & 49.99 & 74.53 & 292.72 \\ 
361.89 & 1.678 & 68.04 & 102.21 & 49.94 & 73.63 & 285.27 \\ 
371.88 & 2.056 & 67.77 & 100.99 & 50.21 & 74.86 & 290.15 \\ 
381.88 & 2.447 & 67.07 & 100.21 & 50.92 & 75.66 & 285.69 \\ 
391.88 & 2.871 & 67.13 & 100.38 & 50.86 & 75.51 & 284.61 \\ 
401.88 & 3.289 & 66.73 & 99.94 & 51.26 & 75.95 & 286.81 \\ 
411.88 & 3.793 & 66.91 & 99.48 & 51.08 & 76.46 & 282.51 \\ 
421.88 & 4.233 & 66.68 & 99.72 & 51.32 & 76.25 & 283.43 \\ 
431.88 & 4.347 & 66.51 & 99.5 & 51.48 & 76.36 & 283.86 \\ 
441.88 & 4.503 & 66.73 & 100.25 & 51.27 & 75.75 & 279.81 \\ 
451.88 & 4.605 & 66.44 & 100.0 & 51.56 & 76.0 & 276.93 \\ 
461.87 & 4.711 & 66.16 & 99.15 & 51.84 & 76.85 & 273.46 \\ 
471.87 & 4.852 & 66.49 & 99.38 & 51.51 & 76.61 & 270.57 \\ 
481.87 & 5.000 & 67.37 & 100.88 & 50.62 & 75.11 & 270.78 \\ 
491.87 & 5.171 & 67.21 & 101.21 & 50.79 & 74.78 & 271.87 \\ 
501.87 & 5.403 & 67.18 & 100.82 & 50.82 & 75.17 & 270.73 \\ 
511.87 & 5.495 & 67.27 & 100.51 & 50.73 & 75.48 & 271.07 \\ 
521.87 & 5.638 & 66.97 & 100.17 & 51.03 & 75.83 & 268.88 \\ 
531.87 & 5.886 & 66.5 & 99.68 & 51.5 & 76.32 & 266.03 \\ 
541.87 & 6.301 & 66.81 & 99.54 & 51.19 & 76.45 & 266.06 \\ 
551.86 & 6.706 & 66.71 & 99.62 & 51.29 & 76.38 & 260.17 \\ 
561.86 & 7.018 & 66.72 & 100.05 & 51.28 & 75.94 & 263.72 \\ 
571.86 & 7.429 & 67.06 & 100.51 & 50.94 & 75.48 & 260.9 \\ 
581.86 & 7.453 & 67.11 & 100.11 & 50.88 & 75.88 & 261.59 \\ 
591.86 & 7.212 & 67.06 & 100.33 & 50.94 & 75.66 & 265.02 \\ 
601.86 & 6.905 & 66.94 & 100.2 & 51.06 & 75.8 & 264.47 \\ 
611.86 & 6.632 & 66.14 & 99.09 & 51.86 & 76.91 & 270.67 \\ 
\end{tabular}
\caption{\label{SnErside} Same as Tab.~\ref{CaCfside} for $^{126}$Sn$+^{168}$Er with the side orientation. }
\end{ruledtabular}
\end{table}

\begin{table}[ht]
\begin{ruledtabular}
\begin{tabular}{ccccccc}
$E_{cm}$ &Contact& $Z_H$& $N_H$&$Z_L$&$N_L$&TKE \\
\hline
306.14 & 0.0 & 67.98 & 99.96 & 50.01 & 76.0 & 304.97 \\ 
316.14 & 0.495 & 67.72 & 99.96 & 50.27 & 75.86 & 294.31 \\ 
326.14 & 1.445 & 67.3 & 99.5 & 50.68 & 76.28 & 268.16 \\ 
336.13 & 1.732 & 66.99 & 99.65 & 51.01 & 76.21 & 277.18 \\ 
346.13 & 2.354 & 66.1 & 98.63 & 51.89 & 77.23 & 259.25 \\ 
356.13 & 2.621 & 65.72 & 97.68 & 52.27 & 78.18 & 254.28 \\ 
366.13 & 2.787 & 66.0 & 98.28 & 51.99 & 77.56 & 246.9 \\ 
376.13 & 3.347 & 63.88 & 94.43 & 54.11 & 81.49 & 237.35 \\ 
386.13 & 3.457 & 61.12 & 92.09 & 56.87 & 83.81 & 237.67 \\ 
396.13 & 3.279 & 62.71 & 91.97 & 55.28 & 83.95 & 236.52 \\ 
406.13 & 2.766 & 65.59 & 96.59 & 52.4 & 79.33 & 251.93 \\ 
416.13 & 2.702 & 65.49 & 96.73 & 52.51 & 79.25 & 251.03 \\ 
426.12 & 2.675 & 65.25 & 96.62 & 52.75 & 79.37 & 250.66 \\ 
436.12 & 2.795 & 65.02 & 97.05 & 52.98 & 78.94 & 243.81 \\ 
446.12 & 3.212 & 66.14 & 99.48 & 51.86 & 76.52 & 229.96 \\ 
456.12 & 3.583 & 61.06 & 90.76 & 56.94 & 85.24 & 234.29 \\ 
466.12 & 3.325 & 59.0 & 88.3 & 59.0 & 87.69 & 234.5 \\ 
476.12 & 3.174 & 60.27 & 89.55 & 57.73 & 86.45 & 239.57 \\ 
486.12 & 3.377 & 59.63 & 89.38 & 58.37 & 86.62 & 225.88 \\ 
496.12 & 3.338 & 60.81 & 90.78 & 57.19 & 85.22 & 230.89 \\ 
506.12 & 3.360 & 59.97 & 89.61 & 58.03 & 86.39 & 231.52 \\ 
516.11 & 3.373 & 59.08 & 88.47 & 58.92 & 87.52 & 232.58 \\ 
526.11 & 3.380 & 61.19 & 90.6 & 56.81 & 85.39 & 234.78 \\ 
536.11 & 3.423 & 61.48 & 91.19 & 56.52 & 84.81 & 242.17 \\ 
546.11 & 3.497 & 61.35 & 91.51 & 56.65 & 84.49 & 242.16 \\ 
556.11 & 3.537 & 62.22 & 93.07 & 55.78 & 82.93 & 240.78 \\ 
566.11 & 3.771 & 63.01 & 94.7 & 54.99 & 81.3 & 231.24 \\ 
576.11 & 3.691 & 64.78 & 96.98 & 53.22 & 79.02 & 232.74 \\
\end{tabular}
\caption{\label{SnErtip} Same as Tab.~\ref{CaCfside} for $^{126}$Sn$+^{168}$Er with the tip orientation. }
\end{ruledtabular}
\end{table}
\clearpage

\bibliography{VU_bibtex_master}

\end{document}